\documentclass[aip,jcp,reprint]{revtex4-1}
\usepackage{amsmath,amsfonts,amssymb}
\usepackage{graphicx}
\usepackage[utf8]{inputenc}
\usepackage{hyperref}

\usepackage[usenames,dvipsnames,svgnames,table]{xcolor}

\begin{document}

\title{Structure diagram of binary Lennard-Jones clusters}

\author{Marko Mravlak}
\affiliation{Research Unit for Physics and Materials Science, Université du Luxembourg, L-1511 Luxembourg, Luxembourg}
\author{Thomas Kister}
\affiliation{INM — Leibniz Institute for New Materials, Campus D2 2, 66123 Saarbrücken, Germany}
\author{Tobias Kraus}
\affiliation{INM — Leibniz Institute for New Materials, Campus D2 2, 66123 Saarbrücken, Germany}
\author{Tanja Schilling}
\affiliation{Research Unit for Physics and Materials Science, Université du Luxembourg, L-1511 Luxembourg, Luxembourg}

\date[Date: ]{\today}

\begin{abstract}
We analyze the structure diagram for binary clusters of Lennard-Jones particles by means of a global optimization approach for a large range of cluster sizes, compositions and interaction energies and present a publicly accessible database of 180,000 minimal energy structures (http://softmattertheory.lu/clusters.html).
We identify a variety of structures such as core-shell clusters, Janus clusters and clusters in which the minority species is located at the vertices of icosahedra.
Such clusters can be synthesized from nanoparticles in agglomeration experiments and used as building blocks in colloidal molecules or crystals.
We discuss the factors that determine the formation of clusters with specific structures.
\end{abstract}


\maketitle

\section{Introduction}

Structured particles are small, regular arrangements of two or more dissimilar components. Such particles have been created, for example, by condensing the vapors of two metals into binary clusters~\cite{Sattler1986}, which had diameters in the nanometer range and uniform structures that minimized their energy~\cite{Baletto2005}.
Colloidal particles can also be assembled into structured clusters, so-called 'supraparticles' that have diameters between nanometers and micrometers. Recently developed self-assembly protocols yield macroscopic quantities of structured supraparticles that are interesting building blocks for nano-structured materials~\cite{mann2009,Arico2005}. Core-shell or Janus particles with anisotropic interactions and valences spontaneously arrange into materials with defined microstructures~\cite{Glotzer2007} or act as surfactants~\cite{Walther2008,Bachinger2011}. It is conceivable that such combinations lead to interesting plasmonic and catalytic behavior, too.

In the case of colloids, the minimal energy configuration is not necessarily always reached. Some assemblies are kinetically trapped and their structures depend on the history of supraparticle formation. (This effect can be exploited to tailor certain supraparticle structures~\cite{Manoharan2006a,Manoharan2003, Chen2014, Park2014}).
However, there are experimental protocols of colloidal assembly that are dominated by energy minimization. For instance, gold nanoparticles in suitably stabilized hexane droplets have been shown to assemble into clusters with structures that are strikingly similar to the global minima of Lennard-Jones clusters~\cite{Lacava2012}. Similarly, iron oxide and silica colloids inside droplets form large, regular nanoparticle clusters that optimize free energy\cite{de2015entropy}.

\begin{figure}
  \centering
  \includegraphics[width=0.48\textwidth]{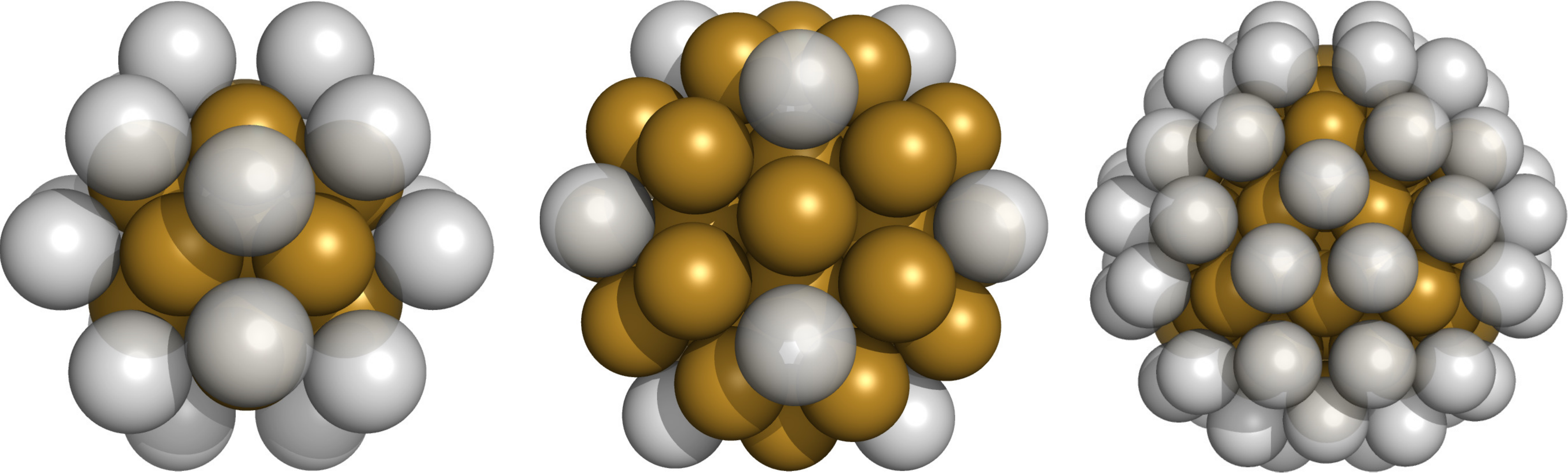}
  \caption{
  Three examples of orientationally symmetric minimal energy clusters of Lennard-Jones particles with the same diameter and with attraction ratio $\epsilon_{BB}=0.50\epsilon_{AA}$
  in which contributions of energetic bonds and dense finite size packing are optimized.
  Left: The ideal icosahedron of A particles in the core is surrounded by 20 B particles at the centers of its triangular surfaces in the minimal energy solution for $N=33$ and $N_B=20$.
  Middle: The ideal icosahedron with valency 12 is a minimal energy solution for $N=55$ and $N_B=12$.
  Right: Minimal energy solution for $N=115$ and $N_B=60$ where B nanoparticles are arranged in 20 triangles lying on top of triangular faces of the central Mackay icosahedron.}
  \label{fig:1}
\end{figure}

So far, no structure diagram has been available to predict which arrangements different particles will assume to minimize their energy. Nanoparticles come with a large range of different sizes and interactions, but existing diagrams are limited to very small subsets. 
The aim of our work is to present a large data base of minimal energy structures for clusters composed of two particle species, and to scan this structure diagram in order to predict parameters for which the ground state has symmetries that are interesting in the context of colloidal molecules and crystals. 
It is known that the structure of minimal energy clusters depends sensitively on the form of the interaction potential~\cite{doye1995effect,DoyeMeyerPRL2005}. As colloidal agglomeration experiments have been shown to produce minimal energy Lennard-Jones cluster structures \cite{Lacava2012} we conclude that the interactions between the nanoparticles in this type of experiment can be modeled by effective Lennard-Jones interactions and that the interactions are strong in comparison to the thermal energy, such that the ground state is formed at room temperature. 

\section{Global optimization}

To find the global energetic minimum of a many-particle system is a difficult mathematical problem. As the energy is a function of a large number of continuous variables which displays many local minima, optimization requires the use of advanced numerical methods.
To minimize energies of heteroparticle systems is even more complicated due to the large number of combinatorial arrangements.
In addition, the less similar the particles are, the more difficult is it to find the global minimum as there are increasingly high energy barriers in the potential energy landscape~\cite{doye2006structure,kolossvary2010global}.
However, the total number of particles in the system remains the most important factor determining the computational effort.

For systems made of just one component the problem has been solved for up to hundreds of particles using unbiased optimization algorithms. In contrast, for multicomponent systems specificities of each system (e.g.~the functional form of the interaction potential and the number of components) need to be taken into account when choosing the optimization strategy.
For the binary Lennard-Jones (BLJ) system, compositional minima for clusters of up to 100 particles with diameter ratios up to 1.3 and one fixed choice of interaction parameters have been computed~\cite{wales2001cambridge, DoyeMeyerPRL2005}. ("Compositional" means that not only the particle positions but also the identities of the particles were varied in order to obtain minimal energies.)
Here, we present a scan of the entire composition diagram, i.e.~a minimal energy structure for each possible choice of composition, for up to 200 particles for multiple sets of interaction parameters. In total, we have computed the global minima for 180,000 different energy landscapes.

\section{Clusters of particles with dissimilar attractions}

We model the self-assembly of a mixture of two different types of spherical particles (A and B) whose attractions differ by a given ratio. Spheres of identical diameter interact via a Lennard-Jones potential
\begin{equation*}
  E = 4\epsilon_{\alpha\beta}\sum_{i<j}\left[\left(\frac{\sigma_{\alpha\beta}}{r_{ij}}\right)^{12}-\left(\frac{\sigma_{\alpha\beta}}{r_{ij}}\right)^{6}\right]
\end{equation*}
where $\alpha$ and $\beta$ label the particle species and we sum over all pairs of particles to obtain the potential energy of a cluster.
In the following, we use $\epsilon_{AA}$ as unit of energy and $\sigma_{AA}$ as unit of length. The free parameters are $\epsilon_{BB}$, $\sigma_{BB}$, $\epsilon_{AB}$, $\sigma_{AB}$, while $N$ and $N_B$ determine the composition of a cluster.
Particles in the cluster are of the same diameter but have different interaction strengths.
We choose different ratios of the interaction constants to describe material combinations of different dissimilarities, $\epsilon_{BB}/\epsilon_{AA}=0.90,0.50\text{ and }0.01$. These values correspond to the ratios of dispersion interactions of gold nanoparticles with those of less strongly interacting materials like silver, copper and polymers, respectively, across a hexane medium~\cite{morrison2002colloidal}.
The remaining parameter was calculated as a geometric mean approximation, $\epsilon_{AB}=\left(\epsilon_{AA}\epsilon_{BB}\right)^{1/2}$, which is also known as the Berthelot combining rule, a standard choice for describing dispersion interactions between two dissimilar materials.
To explore the effect of the combining rule we additionally studied three different values for the inter-species interaction strength at a fixed $\epsilon_{BB}$.

Although we set the parameters to model specific combinations of materials, the results we present are rather general. We tested the stability of several structures against variation of the interaction parameters and found them to be stable over a relatively large range (see fig.~\ref{fig:5}). Thus e.g.~a mixture of metallic and polymeric nanoparticles, as they are commonly used in experiments on colloidal suspensions, would yield the same structures for many different choices of metal.

\section{Methods}

Binary Lennard-Jones clusters have been used as a benchmarking system for global optimization algorithms due to the mathematical complexity they pose to state of the art computational resources~\cite{chill2014benchmarks}.
For multicomponent systems global optimization is especially challenging as it requires besides geometrical optimization of particle positions also a permutational optimization of particle identities.
A common approach to geometrical optimization of clusters is the basin-hopping algorithm which is a Monte-Carlo based method that produces an unbiased walk through a transformed potential energy surface, where in a specified number of Monte-Carlo steps one hopes to reach the lowest minimum.
A transformation into basins of attractions computed by a deterministic local optimization method is employed to facilitate the search on top of a complex energy landscape~\cite{wales1997global}.

Such an algorithm performs well for single particle clusters while for heterogeneous systems additional combinatorial local minimization steps are required to relax cluster configurations with respect to particle types.
In a binary cluster the second part of the algorithm thus aims to find the optimal permutational isomer among possibly $N!/(N_A!+N_B!)$ different \emph{homotops} which differ only by identities assigned to particles in a specific geometrical arrangement.
To find an optimal permutation we use another deterministic scheme based on an iterated local search where a sequence of identity swaps is performed until a termination criterion is met.
This is similar to a graph partitioning heuristic of Kernighan and Lin with the difference that only swaps that produce lower energy are accepted and the next swap is determined by a sequence of approximated flip gains while the iteration terminates if no swap producing a lower energy is found~\cite{SchebarchovWalesJCP2013}.
Every swap of particle identities is followed by a geometrical local optimization which converges to special points in the configuration space called \emph{biminima} - the local minima in both coordinate and permutation space~\cite{SchebarchovWalesPRL2014}.
This variant of the basin hopping algorithm is therefore exploring a sampling domain that is reduced further and consequently outperforms the basic basin-hopping algorithm for binary clusters.
We used the implementation of these algorithms given in the GMIN program~\cite{wales2013gmin}.

\section{Structure diagram}
We analyzed the structure diagram as a function of the cluster size and composition, i.e.~of the number of all particles $N$ and the number of B particles $N_B$.
In contrast to the work of Doye~\cite{DoyeMeyerPRL2005} we are not interested in the compositional global minima, but in the lowest minima at given compositions.
With this we determine the structural behavior that is to be expected for mixtures of spheres with dissimilar attractions in e.g.~confined agglomeration experiments with ligand coated bimetallic nanoparticles where the composition in an individual emulsion droplet is fixed~\cite{Lacava2012}.

\begin{figure}
  \centering
  \scriptsize
  \makeatletter
  \providecommand\color[2][]{
    \GenericError{(gnuplot) \space\space\space\@spaces}{
      Package color not loaded in conjunction with
      terminal option `colourtext'
    }{See the gnuplot documentation for explanation.
    }{Either use 'blacktext' in gnuplot or load the package
      color.sty in LaTeX.}
    \renewcommand\color[2][]{}
  }
  \providecommand\includegraphics[2][]{
    \GenericError{(gnuplot) \space\space\space\@spaces}{
      Package graphicx or graphics not loaded
    }{See the gnuplot documentation for explanation.
    }{The gnuplot epslatex terminal needs graphicx.sty or graphics.sty.}
    \renewcommand\includegraphics[2][]{}
  }
  \providecommand\rotatebox[2]{#2}
  \@ifundefined{ifGPcolor}{
    \newif\ifGPcolor
    \GPcolortrue
  }{}
  \@ifundefined{ifGPblacktext}{
    \newif\ifGPblacktext
    \GPblacktexttrue
  }{}
  
  \let\gplgaddtomacro\g@addto@macro
  
  \gdef\gplbacktext{}
  \gdef\gplfronttext{}
  \makeatother
  \ifGPblacktext
    
    \def\colorrgb#1{}
    \def\colorgray#1{}
  \else
    
    \ifGPcolor
      \def\colorrgb#1{\color[rgb]{#1}}
      \def\colorgray#1{\color[gray]{#1}}
      \expandafter\def\csname LTw\endcsname{\color{white}}
      \expandafter\def\csname LTb\endcsname{\color{black}}
      \expandafter\def\csname LTa\endcsname{\color{black}}
      \expandafter\def\csname LT0\endcsname{\color[rgb]{1,0,0}}
      \expandafter\def\csname LT1\endcsname{\color[rgb]{0,1,0}}
      \expandafter\def\csname LT2\endcsname{\color[rgb]{0,0,1}}
      \expandafter\def\csname LT3\endcsname{\color[rgb]{1,0,1}}
      \expandafter\def\csname LT4\endcsname{\color[rgb]{0,1,1}}
      \expandafter\def\csname LT5\endcsname{\color[rgb]{1,1,0}}
      \expandafter\def\csname LT6\endcsname{\color[rgb]{0,0,0}}
      \expandafter\def\csname LT7\endcsname{\color[rgb]{1,0.3,0}}
      \expandafter\def\csname LT8\endcsname{\color[rgb]{0.5,0.5,0.5}}
    \else
      
      \def\colorrgb#1{\color{black}}
      \def\colorgray#1{\color[gray]{#1}}
      \expandafter\def\csname LTw\endcsname{\color{white}}
      \expandafter\def\csname LTb\endcsname{\color{black}}
      \expandafter\def\csname LTa\endcsname{\color{black}}
      \expandafter\def\csname LT0\endcsname{\color{black}}
      \expandafter\def\csname LT1\endcsname{\color{black}}
      \expandafter\def\csname LT2\endcsname{\color{black}}
      \expandafter\def\csname LT3\endcsname{\color{black}}
      \expandafter\def\csname LT4\endcsname{\color{black}}
      \expandafter\def\csname LT5\endcsname{\color{black}}
      \expandafter\def\csname LT6\endcsname{\color{black}}
      \expandafter\def\csname LT7\endcsname{\color{black}}
      \expandafter\def\csname LT8\endcsname{\color{black}}
    \fi
  \fi
    \setlength{\unitlength}{0.0500bp}
    \ifx\gptboxheight\undefined
      \newlength{\gptboxheight}
      \newlength{\gptboxwidth}
      \newsavebox{\gptboxtext}
    \fi
    \setlength{\fboxrule}{0.5pt}
    \setlength{\fboxsep}{1pt}
  \begin{picture}(4640.00,5000.00)
    \gplgaddtomacro\gplbacktext{
      \csname LTb\endcsname
      \put(526,1001){\makebox(0,0)[r]{\strut{}$0$}}
      \csname LTb\endcsname
      \put(526,1844){\makebox(0,0)[r]{\strut{}$50$}}
      \csname LTb\endcsname
      \put(526,2688){\makebox(0,0)[r]{\strut{}$100$}}
      \csname LTb\endcsname
      \put(526,3531){\makebox(0,0)[r]{\strut{}$150$}}
      \csname LTb\endcsname
      \put(526,4374){\makebox(0,0)[r]{\strut{}$200$}}
      \csname LTb\endcsname
      \put(635,4549){\makebox(0,0){\strut{}$0$}}
      \csname LTb\endcsname
      \put(1478,4549){\makebox(0,0){\strut{}$50$}}
      \csname LTb\endcsname
      \put(2322,4549){\makebox(0,0){\strut{}$100$}}
      \csname LTb\endcsname
      \put(3165,4549){\makebox(0,0){\strut{}$150$}}
      \csname LTb\endcsname
      \put(4008,4549){\makebox(0,0){\strut{}$200$}}
      \csname LTb\endcsname
      \put(4245,200){\makebox(0,0)[l]{\strut{}$d\,[\sigma_{AA}]$}}
      \csname LTb\endcsname
      \put(93,2050){\rotatebox{90}{\makebox(0,0)[l]{\strut{}$N$ $(\epsilon_{BB}=0.01)$}}}
      \csname LTb\endcsname
      \put(1763,4749){\makebox(0,0)[l]{\strut{}$N_B$ $(\epsilon_{BB}=0.01)$}}
    }
    \gplgaddtomacro\gplfronttext{
    }
    \gplgaddtomacro\gplbacktext{
      \csname LTb\endcsname
      \put(719,725){\makebox(0,0){\strut{}$0$}}
      \csname LTb\endcsname
      \put(1558,725){\makebox(0,0){\strut{}$50$}}
      \csname LTb\endcsname
      \put(2396,725){\makebox(0,0){\strut{}$100$}}
      \csname LTb\endcsname
      \put(3235,725){\makebox(0,0){\strut{}$150$}}
      \csname LTb\endcsname
      \put(4073,725){\makebox(0,0){\strut{}$200$}}
      \csname LTb\endcsname
      \put(4182,902){\makebox(0,0)[l]{\strut{}$0$}}
      \csname LTb\endcsname
      \put(4182,1741){\makebox(0,0)[l]{\strut{}$50$}}
      \csname LTb\endcsname
      \put(4182,2580){\makebox(0,0)[l]{\strut{}$100$}}
      \csname LTb\endcsname
      \put(4182,3418){\makebox(0,0)[l]{\strut{}$150$}}
      \csname LTb\endcsname
      \put(4182,4257){\makebox(0,0)[l]{\strut{}$200$}}
      \csname LTb\endcsname
      \put(4245,200){\makebox(0,0)[l]{\strut{}$d\,[\sigma_{AA}]$}}
      \csname LTb\endcsname
      \put(93,2050){\rotatebox{90}{\makebox(0,0)[l]{\strut{}$N$ $(\epsilon_{BB}=0.01)$}}}
      \csname LTb\endcsname
      \put(1763,4749){\makebox(0,0)[l]{\strut{}$N_B$ $(\epsilon_{BB}=0.01)$}}
      \csname LTb\endcsname
      \put(1763,550){\makebox(0,0)[l]{\strut{}$N$ $(\epsilon_{BB}=0.90)$}}
      \csname LTb\endcsname
      \put(4639,3299){\rotatebox{270}{\makebox(0,0)[l]{\strut{}$N_B$ $(\epsilon_{BB}=0.90)$}}}
    }
    \gplgaddtomacro\gplfronttext{
      \csname LTb\endcsname
      \put(603,19){\makebox(0,0){\strut{}$0$}}
      \csname LTb\endcsname
      \put(1467,19){\makebox(0,0){\strut{}$0.2$}}
      \csname LTb\endcsname
      \put(2331,19){\makebox(0,0){\strut{}$0.4$}}
      \csname LTb\endcsname
      \put(3195,19){\makebox(0,0){\strut{}$0.6$}}
      \csname LTb\endcsname
      \put(4059,19){\makebox(0,0){\strut{}$0.8$}}
    }
    \put(0,0){\includegraphics{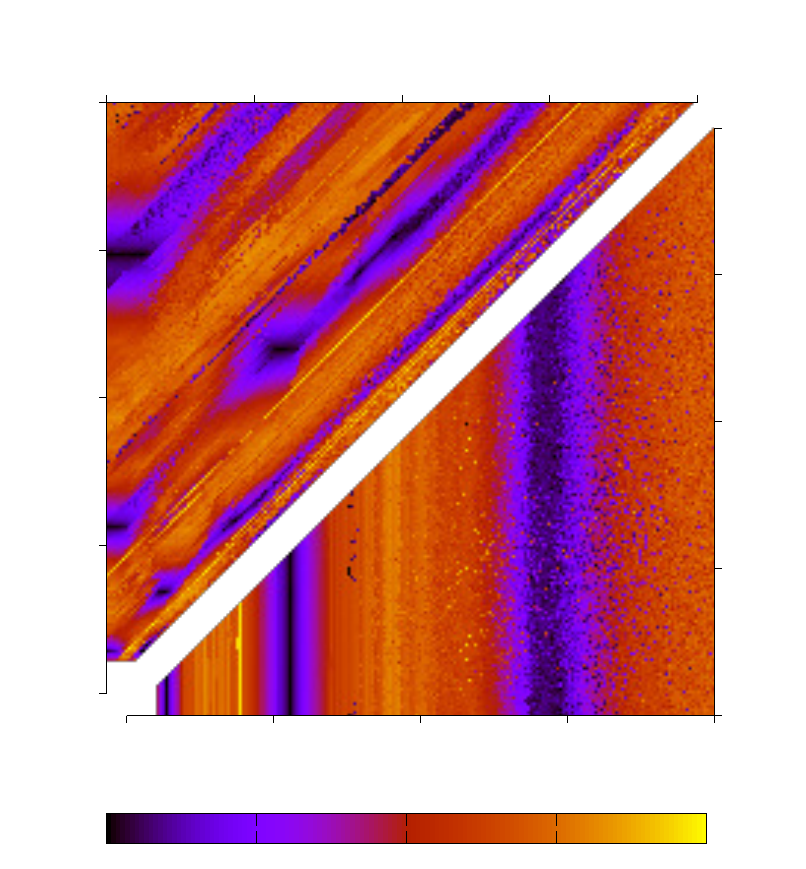}}
    \gplbacktext
    \gplfronttext
  \end{picture}
  \caption{Distance of the innermost particle from the center of the cluster as a function of cluster composition and size for two different material combinations
  with dark regions indicating geometric magic cluster sizes.
  The vertical lines in the lower diagram show that the arrangements close to the complete Mackay icosahedra are not sensitive to composition as opposed to clusters of particles with less similar attractions in the upper diagram, where the diagonal lines indicate a strong dependence on the number of less tightly bound particles.}
  \label{fig:2}
\end{figure}

\begin{figure}
  \centering
  \scriptsize
  \makeatletter
  \providecommand\color[2][]{
    \GenericError{(gnuplot) \space\space\space\@spaces}{
      Package color not loaded in conjunction with
      terminal option `colourtext'
    }{See the gnuplot documentation for explanation.
    }{Either use 'blacktext' in gnuplot or load the package
      color.sty in LaTeX.}
    \renewcommand\color[2][]{}
  }
  \providecommand\includegraphics[2][]{
    \GenericError{(gnuplot) \space\space\space\@spaces}{
      Package graphicx or graphics not loaded
    }{See the gnuplot documentation for explanation.
    }{The gnuplot epslatex terminal needs graphicx.sty or graphics.sty.}
    \renewcommand\includegraphics[2][]{}
  }
  \providecommand\rotatebox[2]{#2}
  \@ifundefined{ifGPcolor}{
    \newif\ifGPcolor
    \GPcolortrue
  }{}
  \@ifundefined{ifGPblacktext}{
    \newif\ifGPblacktext
    \GPblacktexttrue
  }{}
  
  \let\gplgaddtomacro\g@addto@macro
  
  \gdef\gplbacktext{}
  \gdef\gplfronttext{}
  \makeatother
  \ifGPblacktext
    
    \def\colorrgb#1{}
    \def\colorgray#1{}
  \else
    
    \ifGPcolor
      \def\colorrgb#1{\color[rgb]{#1}}
      \def\colorgray#1{\color[gray]{#1}}
      \expandafter\def\csname LTw\endcsname{\color{white}}
      \expandafter\def\csname LTb\endcsname{\color{black}}
      \expandafter\def\csname LTa\endcsname{\color{black}}
      \expandafter\def\csname LT0\endcsname{\color[rgb]{1,0,0}}
      \expandafter\def\csname LT1\endcsname{\color[rgb]{0,1,0}}
      \expandafter\def\csname LT2\endcsname{\color[rgb]{0,0,1}}
      \expandafter\def\csname LT3\endcsname{\color[rgb]{1,0,1}}
      \expandafter\def\csname LT4\endcsname{\color[rgb]{0,1,1}}
      \expandafter\def\csname LT5\endcsname{\color[rgb]{1,1,0}}
      \expandafter\def\csname LT6\endcsname{\color[rgb]{0,0,0}}
      \expandafter\def\csname LT7\endcsname{\color[rgb]{1,0.3,0}}
      \expandafter\def\csname LT8\endcsname{\color[rgb]{0.5,0.5,0.5}}
    \else
      
      \def\colorrgb#1{\color{black}}
      \def\colorgray#1{\color[gray]{#1}}
      \expandafter\def\csname LTw\endcsname{\color{white}}
      \expandafter\def\csname LTb\endcsname{\color{black}}
      \expandafter\def\csname LTa\endcsname{\color{black}}
      \expandafter\def\csname LT0\endcsname{\color{black}}
      \expandafter\def\csname LT1\endcsname{\color{black}}
      \expandafter\def\csname LT2\endcsname{\color{black}}
      \expandafter\def\csname LT3\endcsname{\color{black}}
      \expandafter\def\csname LT4\endcsname{\color{black}}
      \expandafter\def\csname LT5\endcsname{\color{black}}
      \expandafter\def\csname LT6\endcsname{\color{black}}
      \expandafter\def\csname LT7\endcsname{\color{black}}
      \expandafter\def\csname LT8\endcsname{\color{black}}
    \fi
  \fi
    \setlength{\unitlength}{0.0500bp}
    \ifx\gptboxheight\undefined
      \newlength{\gptboxheight}
      \newlength{\gptboxwidth}
      \newsavebox{\gptboxtext}
    \fi
    \setlength{\fboxrule}{0.5pt}
    \setlength{\fboxsep}{1pt}
  \begin{picture}(4640.00,5000.00)
    \gplgaddtomacro\gplbacktext{
      \csname LTb\endcsname
      \put(526,1001){\makebox(0,0)[r]{\strut{}$0$}}
      \csname LTb\endcsname
      \put(526,1844){\makebox(0,0)[r]{\strut{}$50$}}
      \csname LTb\endcsname
      \put(526,2688){\makebox(0,0)[r]{\strut{}$100$}}
      \csname LTb\endcsname
      \put(526,3531){\makebox(0,0)[r]{\strut{}$150$}}
      \csname LTb\endcsname
      \put(526,4374){\makebox(0,0)[r]{\strut{}$200$}}
      \csname LTb\endcsname
      \put(635,4549){\makebox(0,0){\strut{}$0$}}
      \csname LTb\endcsname
      \put(1478,4549){\makebox(0,0){\strut{}$50$}}
      \csname LTb\endcsname
      \put(2322,4549){\makebox(0,0){\strut{}$100$}}
      \csname LTb\endcsname
      \put(3165,4549){\makebox(0,0){\strut{}$150$}}
      \csname LTb\endcsname
      \put(4008,4549){\makebox(0,0){\strut{}$200$}}
      \csname LTb\endcsname
      \put(93,2050){\rotatebox{90}{\makebox(0,0)[l]{\strut{}$N$ $(\epsilon_{BB}=0.01)$}}}
      \csname LTb\endcsname
      \put(1763,4749){\makebox(0,0)[l]{\strut{}$N_B$ $(\epsilon_{BB}=0.01)$}}
    }
    \gplgaddtomacro\gplfronttext{
    }
    \gplgaddtomacro\gplbacktext{
      \csname LTb\endcsname
      \put(719,725){\makebox(0,0){\strut{}$0$}}
      \csname LTb\endcsname
      \put(1558,725){\makebox(0,0){\strut{}$50$}}
      \csname LTb\endcsname
      \put(2396,725){\makebox(0,0){\strut{}$100$}}
      \csname LTb\endcsname
      \put(3235,725){\makebox(0,0){\strut{}$150$}}
      \csname LTb\endcsname
      \put(4073,725){\makebox(0,0){\strut{}$200$}}
      \csname LTb\endcsname
      \put(4182,902){\makebox(0,0)[l]{\strut{}$0$}}
      \csname LTb\endcsname
      \put(4182,1741){\makebox(0,0)[l]{\strut{}$50$}}
      \csname LTb\endcsname
      \put(4182,2580){\makebox(0,0)[l]{\strut{}$100$}}
      \csname LTb\endcsname
      \put(4182,3418){\makebox(0,0)[l]{\strut{}$150$}}
      \csname LTb\endcsname
      \put(4182,4257){\makebox(0,0)[l]{\strut{}$200$}}
      \csname LTb\endcsname
      \put(93,2050){\rotatebox{90}{\makebox(0,0)[l]{\strut{}$N$ $(\epsilon_{BB}=0.01)$}}}
      \csname LTb\endcsname
      \put(1763,4749){\makebox(0,0)[l]{\strut{}$N_B$ $(\epsilon_{BB}=0.01)$}}
      \csname LTb\endcsname
      \put(1763,550){\makebox(0,0)[l]{\strut{}$N$ $(\epsilon_{BB}=0.90)$}}
      \csname LTb\endcsname
      \put(4639,3299){\rotatebox{270}{\makebox(0,0)[l]{\strut{}$N_B$ $(\epsilon_{BB}=0.90)$}}}
    }
    \gplgaddtomacro\gplfronttext{
      \csname LTb\endcsname
      \put(879,19){\makebox(0,0){\strut{}fcc}}
      \csname LTb\endcsname
      \put(1467,19){\makebox(0,0){\strut{}hcp}}
      \csname LTb\endcsname
      \put(2054,19){\makebox(0,0){\strut{}sc}}
      \csname LTb\endcsname
      \put(2642,19){\makebox(0,0){\strut{}bcc}}
      \csname LTb\endcsname
      \put(3222,19){\makebox(0,0){\strut{}liq}}
      \csname LTb\endcsname
      \put(3796,19){\makebox(0,0){\strut{}ih}}
    }
    \put(0,0){\includegraphics{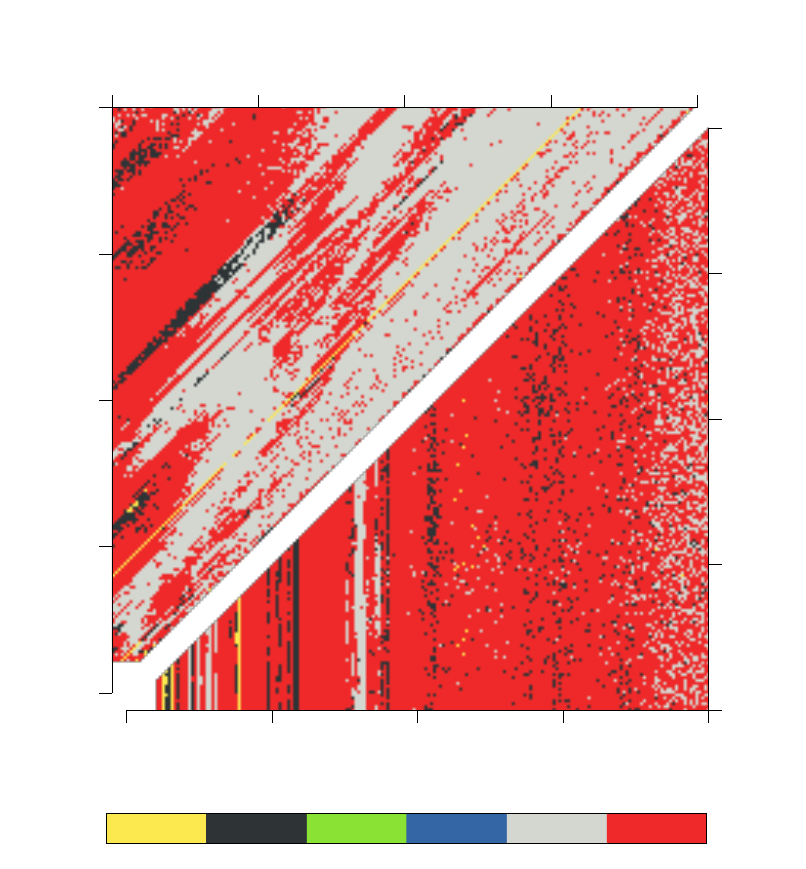}}
    \gplbacktext
    \gplfronttext
  \end{picture}
  \caption{Classification of clusters with different sizes and compositions according to the bond order parameters $q_4$, $q_6$, $w_4$ and $w_6$ averaged over all the particles in the cluster for two different material combinations.}
  \label{fig:3}
\end{figure}

\begin{figure}[b]
  \centering
    \scriptsize
  \makeatletter
  \providecommand\color[2][]{
    \GenericError{(gnuplot) \space\space\space\@spaces}{
      Package color not loaded in conjunction with
      terminal option `colourtext'
    }{See the gnuplot documentation for explanation.
    }{Either use 'blacktext' in gnuplot or load the package
      color.sty in LaTeX.}
    \renewcommand\color[2][]{}
  }
  \providecommand\includegraphics[2][]{
    \GenericError{(gnuplot) \space\space\space\@spaces}{
      Package graphicx or graphics not loaded
    }{See the gnuplot documentation for explanation.
    }{The gnuplot epslatex terminal needs graphicx.sty or graphics.sty.}
    \renewcommand\includegraphics[2][]{}
  }
  \providecommand\rotatebox[2]{#2}
  \@ifundefined{ifGPcolor}{
    \newif\ifGPcolor
    \GPcolortrue
  }{}
  \@ifundefined{ifGPblacktext}{
    \newif\ifGPblacktext
    \GPblacktexttrue
  }{}
  
  \let\gplgaddtomacro\g@addto@macro
  
  \gdef\gplbacktext{}
  \gdef\gplfronttext{}
  \makeatother
  \ifGPblacktext
    
    \def\colorrgb#1{}
    \def\colorgray#1{}
  \else
    
    \ifGPcolor
      \def\colorrgb#1{\color[rgb]{#1}}
      \def\colorgray#1{\color[gray]{#1}}
      \expandafter\def\csname LTw\endcsname{\color{white}}
      \expandafter\def\csname LTb\endcsname{\color{black}}
      \expandafter\def\csname LTa\endcsname{\color{black}}
      \expandafter\def\csname LT0\endcsname{\color[rgb]{1,0,0}}
      \expandafter\def\csname LT1\endcsname{\color[rgb]{0,1,0}}
      \expandafter\def\csname LT2\endcsname{\color[rgb]{0,0,1}}
      \expandafter\def\csname LT3\endcsname{\color[rgb]{1,0,1}}
      \expandafter\def\csname LT4\endcsname{\color[rgb]{0,1,1}}
      \expandafter\def\csname LT5\endcsname{\color[rgb]{1,1,0}}
      \expandafter\def\csname LT6\endcsname{\color[rgb]{0,0,0}}
      \expandafter\def\csname LT7\endcsname{\color[rgb]{1,0.3,0}}
      \expandafter\def\csname LT8\endcsname{\color[rgb]{0.5,0.5,0.5}}
    \else
      
      \def\colorrgb#1{\color{black}}
      \def\colorgray#1{\color[gray]{#1}}
      \expandafter\def\csname LTw\endcsname{\color{white}}
      \expandafter\def\csname LTb\endcsname{\color{black}}
      \expandafter\def\csname LTa\endcsname{\color{black}}
      \expandafter\def\csname LT0\endcsname{\color{black}}
      \expandafter\def\csname LT1\endcsname{\color{black}}
      \expandafter\def\csname LT2\endcsname{\color{black}}
      \expandafter\def\csname LT3\endcsname{\color{black}}
      \expandafter\def\csname LT4\endcsname{\color{black}}
      \expandafter\def\csname LT5\endcsname{\color{black}}
      \expandafter\def\csname LT6\endcsname{\color{black}}
      \expandafter\def\csname LT7\endcsname{\color{black}}
      \expandafter\def\csname LT8\endcsname{\color{black}}
    \fi
  \fi
    \setlength{\unitlength}{0.0500bp}
    \ifx\gptboxheight\undefined
      \newlength{\gptboxheight}
      \newlength{\gptboxwidth}
      \newsavebox{\gptboxtext}
    \fi
    \setlength{\fboxrule}{0.5pt}
    \setlength{\fboxsep}{1pt}
  \begin{picture}(4640.00,5000.00)
    \gplgaddtomacro\gplbacktext{
      \csname LTb\endcsname
      \put(526,1001){\makebox(0,0)[r]{\strut{}$0$}}
      \csname LTb\endcsname
      \put(526,1844){\makebox(0,0)[r]{\strut{}$50$}}
      \csname LTb\endcsname
      \put(526,2688){\makebox(0,0)[r]{\strut{}$100$}}
      \csname LTb\endcsname
      \put(526,3531){\makebox(0,0)[r]{\strut{}$150$}}
      \csname LTb\endcsname
      \put(526,4374){\makebox(0,0)[r]{\strut{}$200$}}
      \csname LTb\endcsname
      \put(635,4549){\makebox(0,0){\strut{}$0$}}
      \csname LTb\endcsname
      \put(1478,4549){\makebox(0,0){\strut{}$50$}}
      \csname LTb\endcsname
      \put(2322,4549){\makebox(0,0){\strut{}$100$}}
      \csname LTb\endcsname
      \put(3165,4549){\makebox(0,0){\strut{}$150$}}
      \csname LTb\endcsname
      \put(4008,4549){\makebox(0,0){\strut{}$200$}}
      \csname LTb\endcsname
      \put(93,2050){\rotatebox{90}{\makebox(0,0)[l]{\strut{}$N$ $(\epsilon_{BB}=0.01)$}}}
      \csname LTb\endcsname
      \put(1763,4749){\makebox(0,0)[l]{\strut{}$N_B$ $(\epsilon_{BB}=0.01)$}}
    }
    \gplgaddtomacro\gplfronttext{
    }
    \gplgaddtomacro\gplbacktext{
      \csname LTb\endcsname
      \put(719,725){\makebox(0,0){\strut{}$0$}}
      \csname LTb\endcsname
      \put(1558,725){\makebox(0,0){\strut{}$50$}}
      \csname LTb\endcsname
      \put(2396,725){\makebox(0,0){\strut{}$100$}}
      \csname LTb\endcsname
      \put(3235,725){\makebox(0,0){\strut{}$150$}}
      \csname LTb\endcsname
      \put(4073,725){\makebox(0,0){\strut{}$200$}}
      \csname LTb\endcsname
      \put(4182,902){\makebox(0,0)[l]{\strut{}$0$}}
      \csname LTb\endcsname
      \put(4182,1741){\makebox(0,0)[l]{\strut{}$50$}}
      \csname LTb\endcsname
      \put(4182,2580){\makebox(0,0)[l]{\strut{}$100$}}
      \csname LTb\endcsname
      \put(4182,3418){\makebox(0,0)[l]{\strut{}$150$}}
      \csname LTb\endcsname
      \put(4182,4257){\makebox(0,0)[l]{\strut{}$200$}}
      \csname LTb\endcsname
      \put(93,2050){\rotatebox{90}{\makebox(0,0)[l]{\strut{}$N$ $(\epsilon_{BB}=0.01)$}}}
      \csname LTb\endcsname
      \put(1763,4749){\makebox(0,0)[l]{\strut{}$N_B$ $(\epsilon_{BB}=0.01)$}}
      \csname LTb\endcsname
      \put(1763,550){\makebox(0,0)[l]{\strut{}$N$ $(\epsilon_{BB}=0.90)$}}
      \csname LTb\endcsname
      \put(4639,3299){\rotatebox{270}{\makebox(0,0)[l]{\strut{}$N_B$ $(\epsilon_{BB}=0.90)$}}}
    }
    \gplgaddtomacro\gplfronttext{
      \csname LTb\endcsname
      \put(603,19){\makebox(0,0){\strut{}$0$}}
      \csname LTb\endcsname
      \put(1179,19){\makebox(0,0){\strut{}$0.1$}}
      \csname LTb\endcsname
      \put(1755,19){\makebox(0,0){\strut{}$0.2$}}
      \csname LTb\endcsname
      \put(2331,19){\makebox(0,0){\strut{}$0.3$}}
      \csname LTb\endcsname
      \put(2907,19){\makebox(0,0){\strut{}$0.4$}}
      \csname LTb\endcsname
      \put(3482,19){\makebox(0,0){\strut{}$0.5$}}
      \csname LTb\endcsname
      \put(4058,19){\makebox(0,0){\strut{}$0.6$}}
    }
    \put(0,0){\includegraphics{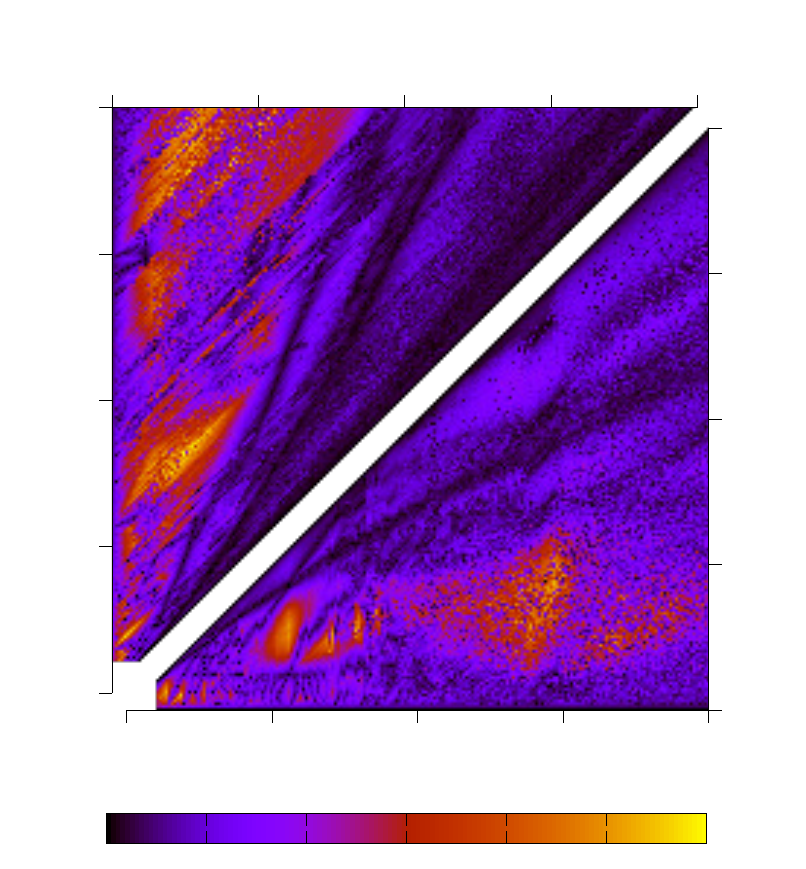}}
    \gplbacktext
    \gplfronttext
  \end{picture}
  \caption{Magnitude of the cluster dipole moment normalized to the number of particles as a function of cluster composition and size for two different material combinations. When computing the dipole moment we assigned charge $+1$ to particles of type~A and charge $-1$ to particles of type~B.}
  \label{fig:4}
\end{figure}

In figures~\ref{fig:2},~\ref{fig:3} and~\ref{fig:4} we present diagrams of several measures that characterize the structure of the clusters for two different material dissimilarities. Data for i.e. $\epsilon_{BB} = 0.90$, $\epsilon_{AB} = 0.95$ is shown in the lower triangle and $\epsilon_{BB} = 0.01$, $\epsilon_{AB} = 0.10$ in the upper triangle. Every speck corresponds to one energy minimum.
For all cluster sizes $N$ and compositions $N_B$ we find a core-shell separated structure with B particles on the outside. This is expected as the A particles attract each other more strongly than the B particles. Clusters with directional symmetries can thus only occur for large number ratios between A and B particles, where the A particles form a core that is covered by a suitable, smaller number of less strongly bound B particles distributed to optimize the mutual energetic interactions.
Diagrams for additional measures and parameter choices are presented in the Supplemental Material (SM)~\cite{SupplementalMaterial} where we also show results for varying mixing attractions of different particle species.
A database containing more than 180,000 minimal energy configurations of the resulting clusters can be visualized and downloaded using our web application~\cite{clustersweb}.  

Figure~\ref{fig:2} shows the minimal distances from the center of the cluster, $d_{\rm min}=\min\left\|\boldsymbol{r}_i-\boldsymbol{r}_c \right\|$. 
At certain numbers, there are clusters with one particle in the center (dark spots), a feature that is characteristic for Mackay icosahedra. We can clearly see this feature around the geometric magic sizes 13, 55 and 147 where complete icosahedra are formed. For large $\epsilon_{BB}$ the Mackay structure is nearly independent of composition (vertical dark stripes) while for small $\epsilon_{BB}$ it is stable for constant $N_A$ (diagonal dark stripes).
Varying the composition in clusters of particles with more similar energetic contributions disturbs the clusters' structures much less significantly than changing the ratio of particles with small energetic contributions where the number of strongly bound particles alone determines the structure. 
In addition, there are some other isolated regions in the structure diagram with a central particle. Here, typically an ideal icosahedral core is covered with a shell of particles of the other species. These structures are particularly interesting as they offer symmetric structures with specific valences where particles with smaller energetic contributions are e.g. in the vertices of icosahedron where the number of nearest neighbors is smaller (see example snapshots in fig.~\ref{fig:1}.). The particles on the corners could be functionalized to produce building blocks of colloidal molecules as desired in the equilibrium self-assembly of complex structures~\cite{grunwald2014patterns}.

We analyze the crystalline structure in terms of Steinhardt bond-orientational order parameters $q_4$ and $q_6$~\cite{steinhardt1983bond}. By comparing values averaged over all particles in the cluster to the values obtained for several known crystals and complete icosahedra we classify clusters according to the closest match and observe that icosahedral features largely prevail~\cite{Wang2005} (which is also confirmed by visually inspecting the resulting clusters).
An interesting feature are the vertical stripes in the lower triangle of fig.~\ref{fig:3} which imply structural features that are independent of composition, i.e.~close packing in space is more important than the optimization of energetic bonds.

\begin{figure}
  \centering
  \includegraphics[width=0.48\textwidth]{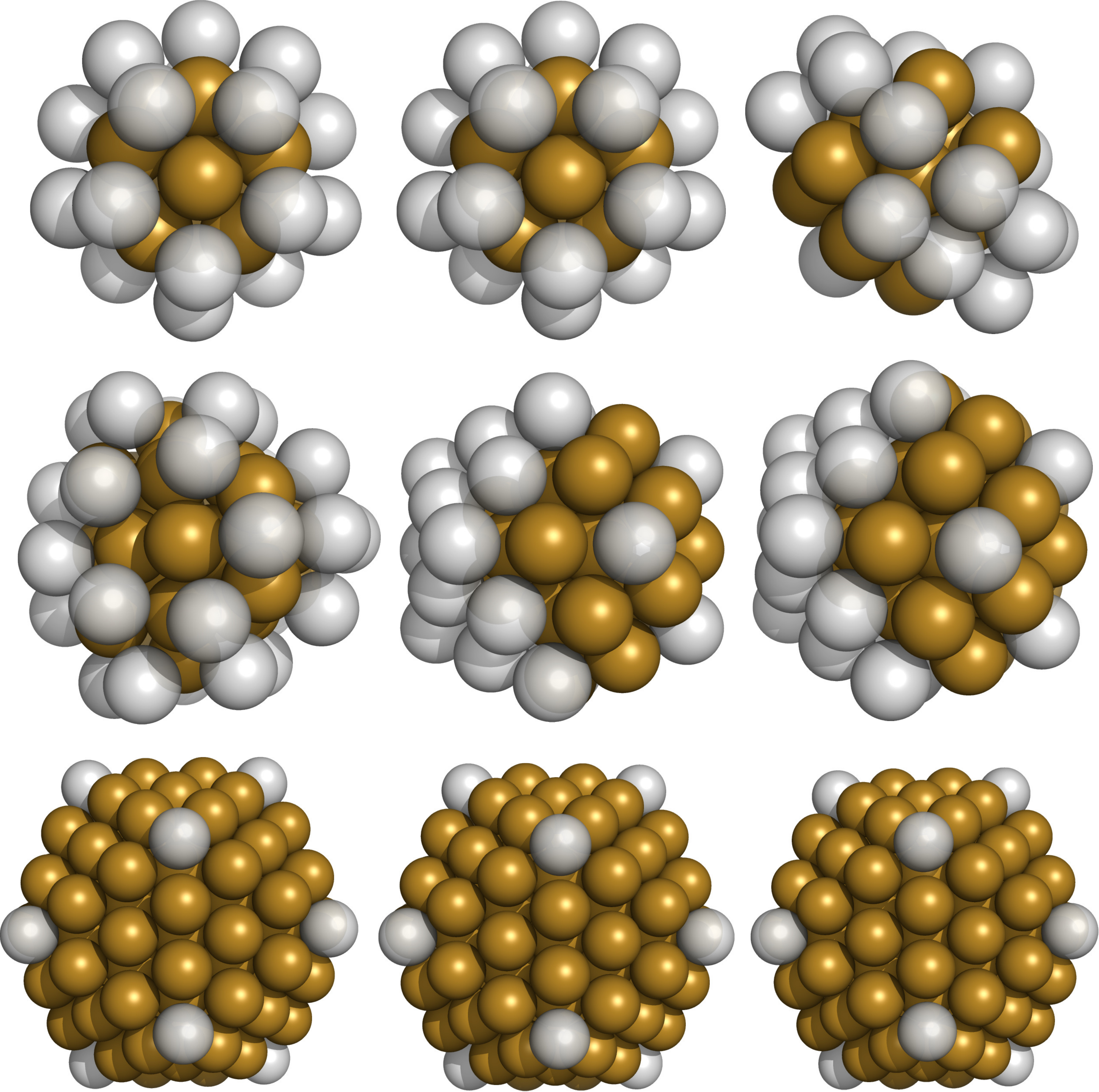}
  \caption{
  Examples of symmetric clusters that show little or no sensitivity to changing the relative attraction strength $\epsilon_{BB}$ of species B particles.
  From left to right: $\epsilon_{BB}=0.01,\;0.5,\;0.9$. From top to bottom: $N=33,\;55,\;147$ and $N_B=20,\;28,\;12$.
  }
  \label{fig:5}
\end{figure}

To quantify the polarity in the distribution of particle species in the clusters we calculate their dipole moments by assigning a charge $+1$ to particles of type~A and a charge $-1$ to particles of type~B (figure~\ref{fig:4}). (Note that this is purely a measure to quantify the spatial separation, we do not assume that there are electric charges.)
If the Berthelot rule is applied to the binary Lennard-Jones interaction, it favors mixing of different particle types more than the aggregation of the less attractive particles,
nevertheless the smaller number of neighbors on the surface of the clusters does contribute and thus the clusters demix into a core of tightly bound particles and a shell with symmetric arrangements of weakly bound particles.
Due to this core-shell structure we do not observe large values of dipole moments.
However, there are also certain regions with larger dipole moments which
correspond to Janus-like phase separation of the particles on the shell of the clusters (see e.g.~fig.~\ref{fig:5} middle row right column).
This is typical for clusters based on complete icosahedra
where either a number of B type particles that is larger than the number of vertices constitutes the outer shell of the icosahedron
or where a complete icosahedron in the core is covered with less attractive particles that arrange on one side of the surface.
Janus clusters with much larger dipole moments that are internally phase separated are observed for mixtures of two dissimilar and less compatible particle types as presented in SM.
Such a model describes the case of disfavored mixing, for example when two less cross compatible ligands are attached to different nanoparticle types.

For the cluster made of 55 particles in fig~\ref{fig:5} we also observe the transition from an isotropic to anisotropic core-shell structure with increasing $\epsilon_{BB}$.
At low $\epsilon_{BB}$ negligible energetic contributions from weakly bound B type particles do not manage to disturb the structure of the strongly bound particles in the core and thus they get isotropically distributed on the shell to maximize the number of energetically more favorable A-B contacts.
At high $\epsilon_{BB}$ both particle types form a complete icosahedron together and due to the number of B type particles being larger than the number of vertices on the surface of the icosahedra they get anisotropically distributed allowing for more A-A contacts on the surface of the cluster.

\begin{figure}
  \centering
  % GNUPLOT: LaTeX picture with Postscript
\begingroup
  \makeatletter
  \providecommand\color[2][]{%
    \GenericError{(gnuplot) \space\space\space\@spaces}{%
      Package color not loaded in conjunction with
      terminal option `colourtext'%
    }{See the gnuplot documentation for explanation.%
    }{Either use 'blacktext' in gnuplot or load the package
      color.sty in LaTeX.}%
    \renewcommand\color[2][]{}%
  }%
  \providecommand\includegraphics[2][]{%
    \GenericError{(gnuplot) \space\space\space\@spaces}{%
      Package graphicx or graphics not loaded%
    }{See the gnuplot documentation for explanation.%
    }{The gnuplot epslatex terminal needs graphicx.sty or graphics.sty.}%
    \renewcommand\includegraphics[2][]{}%
  }%
  \providecommand\rotatebox[2]{#2}%
  \@ifundefined{ifGPcolor}{%
    \newif\ifGPcolor
    \GPcolortrue
  }{}%
  \@ifundefined{ifGPblacktext}{%
    \newif\ifGPblacktext
    \GPblacktexttrue
  }{}%
  % define a \g@addto@macro without @ in the name:
  \let\gplgaddtomacro\g@addto@macro
  % define empty templates for all commands taking text:
  \gdef\gplbacktext{}%
  \gdef\gplfronttext{}%
  \makeatother
  \ifGPblacktext
    % no textcolor at all
    \def\colorrgb#1{}%
    \def\colorgray#1{}%
  \else
    % gray or color?
    \ifGPcolor
      \def\colorrgb#1{\color[rgb]{#1}}%
      \def\colorgray#1{\color[gray]{#1}}%
      \expandafter\def\csname LTw\endcsname{\color{white}}%
      \expandafter\def\csname LTb\endcsname{\color{black}}%
      \expandafter\def\csname LTa\endcsname{\color{black}}%
      \expandafter\def\csname LT0\endcsname{\color[rgb]{1,0,0}}%
      \expandafter\def\csname LT1\endcsname{\color[rgb]{0,1,0}}%
      \expandafter\def\csname LT2\endcsname{\color[rgb]{0,0,1}}%
      \expandafter\def\csname LT3\endcsname{\color[rgb]{1,0,1}}%
      \expandafter\def\csname LT4\endcsname{\color[rgb]{0,1,1}}%
      \expandafter\def\csname LT5\endcsname{\color[rgb]{1,1,0}}%
      \expandafter\def\csname LT6\endcsname{\color[rgb]{0,0,0}}%
      \expandafter\def\csname LT7\endcsname{\color[rgb]{1,0.3,0}}%
      \expandafter\def\csname LT8\endcsname{\color[rgb]{0.5,0.5,0.5}}%
    \else
      % gray
      \def\colorrgb#1{\color{black}}%
      \def\colorgray#1{\color[gray]{#1}}%
      \expandafter\def\csname LTw\endcsname{\color{white}}%
      \expandafter\def\csname LTb\endcsname{\color{black}}%
      \expandafter\def\csname LTa\endcsname{\color{black}}%
      \expandafter\def\csname LT0\endcsname{\color{black}}%
      \expandafter\def\csname LT1\endcsname{\color{black}}%
      \expandafter\def\csname LT2\endcsname{\color{black}}%
      \expandafter\def\csname LT3\endcsname{\color{black}}%
      \expandafter\def\csname LT4\endcsname{\color{black}}%
      \expandafter\def\csname LT5\endcsname{\color{black}}%
      \expandafter\def\csname LT6\endcsname{\color{black}}%
      \expandafter\def\csname LT7\endcsname{\color{black}}%
      \expandafter\def\csname LT8\endcsname{\color{black}}%
    \fi
  \fi
    \setlength{\unitlength}{0.0500bp}%
    \ifx\gptboxheight\undefined%
      \newlength{\gptboxheight}%
      \newlength{\gptboxwidth}%
      \newsavebox{\gptboxtext}%
    \fi%
    \setlength{\fboxrule}{0.5pt}%
    \setlength{\fboxsep}{1pt}%
\begin{picture}(4640.00,5000.00)%
    \gplgaddtomacro\gplbacktext{%
      \csname LTb\endcsname%
      \put(526,1001){\makebox(0,0)[r]{\strut{}$0$}}%
      \csname LTb\endcsname%
      \put(526,1844){\makebox(0,0)[r]{\strut{}$50$}}%
      \csname LTb\endcsname%
      \put(526,2688){\makebox(0,0)[r]{\strut{}$100$}}%
      \csname LTb\endcsname%
      \put(526,3531){\makebox(0,0)[r]{\strut{}$150$}}%
      \csname LTb\endcsname%
      \put(526,4374){\makebox(0,0)[r]{\strut{}$200$}}%
      \csname LTb\endcsname%
      \put(635,4549){\makebox(0,0){\strut{}$0$}}%
      \csname LTb\endcsname%
      \put(1478,4549){\makebox(0,0){\strut{}$50$}}%
      \csname LTb\endcsname%
      \put(2322,4549){\makebox(0,0){\strut{}$100$}}%
      \csname LTb\endcsname%
      \put(3165,4549){\makebox(0,0){\strut{}$150$}}%
      \csname LTb\endcsname%
      \put(4008,4549){\makebox(0,0){\strut{}$200$}}%
      \csname LTb\endcsname%
      \put(4245,200){\makebox(0,0)[l]{\strut{}$d\,[\sigma_{AA}]$}}%
      \csname LTb\endcsname%
      \put(93,2050){\rotatebox{90}{\makebox(0,0)[l]{\strut{}$N$ $(\epsilon_{BB}=0.10)$}}}%
      \csname LTb\endcsname%
      \put(1763,4749){\makebox(0,0)[l]{\strut{}$N_B$ $(\epsilon_{BB}=0.10)$}}%
    }%
    \gplgaddtomacro\gplfronttext{%
    }%
    \gplgaddtomacro\gplbacktext{%
      \csname LTb\endcsname%
      \put(719,725){\makebox(0,0){\strut{}$0$}}%
      \csname LTb\endcsname%
      \put(1558,725){\makebox(0,0){\strut{}$50$}}%
      \csname LTb\endcsname%
      \put(2396,725){\makebox(0,0){\strut{}$100$}}%
      \csname LTb\endcsname%
      \put(3235,725){\makebox(0,0){\strut{}$150$}}%
      \csname LTb\endcsname%
      \put(4073,725){\makebox(0,0){\strut{}$200$}}%
      \csname LTb\endcsname%
      \put(4182,902){\makebox(0,0)[l]{\strut{}$0$}}%
      \csname LTb\endcsname%
      \put(4182,1741){\makebox(0,0)[l]{\strut{}$50$}}%
      \csname LTb\endcsname%
      \put(4182,2580){\makebox(0,0)[l]{\strut{}$100$}}%
      \csname LTb\endcsname%
      \put(4182,3418){\makebox(0,0)[l]{\strut{}$150$}}%
      \csname LTb\endcsname%
      \put(4182,4257){\makebox(0,0)[l]{\strut{}$200$}}%
      \csname LTb\endcsname%
      \put(4245,200){\makebox(0,0)[l]{\strut{}$d\,[\sigma_{AA}]$}}%
      \csname LTb\endcsname%
      \put(93,2050){\rotatebox{90}{\makebox(0,0)[l]{\strut{}$N$ $(\epsilon_{BB}=0.10)$}}}%
      \csname LTb\endcsname%
      \put(1763,4749){\makebox(0,0)[l]{\strut{}$N_B$ $(\epsilon_{BB}=0.10)$}}%
      \csname LTb\endcsname%
      \put(1763,550){\makebox(0,0)[l]{\strut{}$N$ $(\epsilon_{BB}=0.50)$}}%
      \csname LTb\endcsname%
      \put(4639,3299){\rotatebox{270}{\makebox(0,0)[l]{\strut{}$N_B$ $(\epsilon_{BB}=0.50)$}}}%
    }%
    \gplgaddtomacro\gplfronttext{%
      \csname LTb\endcsname%
      \put(603,19){\makebox(0,0){\strut{}$0$}}%
      \csname LTb\endcsname%
      \put(1467,19){\makebox(0,0){\strut{}$0.2$}}%
      \csname LTb\endcsname%
      \put(2331,19){\makebox(0,0){\strut{}$0.4$}}%
      \csname LTb\endcsname%
      \put(3195,19){\makebox(0,0){\strut{}$0.6$}}%
      \csname LTb\endcsname%
      \put(4059,19){\makebox(0,0){\strut{}$0.8$}}%
    }%
    \put(0,0){\includegraphics{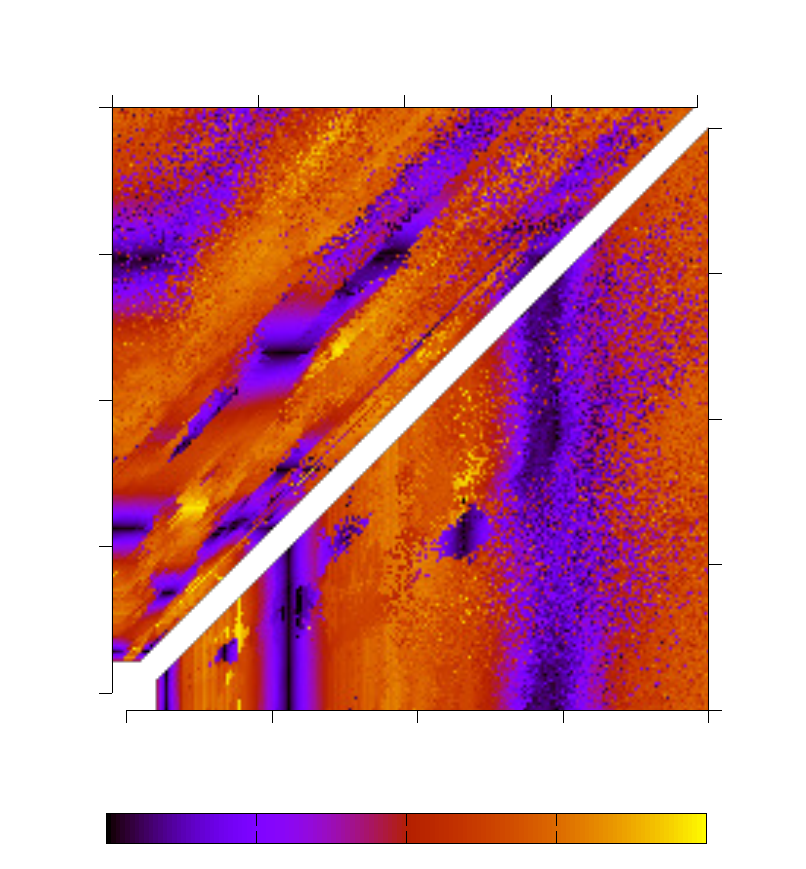}}%
    \gplbacktext
    \gplfronttext
  \end{picture}%
\endgroup
  \caption{Distance of the innermost particle from the center of the cluster as a function of cluster composition and size for two different material combinations with $\epsilon=0.10$ and $\epsilon=0.50$.
  }
  \label{fig:6}
\end{figure}
  
The structures of most symmetric clusters are not affected by the relative attraction strength over a large range of values of $\epsilon_{BB}$ (see e.g.~fig.~\ref{fig:5}).
This indicates a very high stability of the icosahedral structural motifs for binary mixtures of particles with almost arbitrarily dissimilar dispersion attractions.
We can further explore this in fig.~\ref{fig:6} where the innermost particle distances are shown for two material combinations with more moderate values of attraction parameters.
In comparison to the previously analyzed parameters (fig~\ref{fig:2}) we see that the general features are only slightly distorted.
In the lower diagram for $\epsilon=0.50$ we also see that new regions with magic compositions~\cite{Baletto2005} appeared. They contain the structures presented in fig~\ref{fig:1} where a complete Mackay icosahedron in the core is covered by a symmetric shell of less attractive particles.

\section{Conclusions}

In summary, we have computed the minimal energy configurations of binary Lennard-Jones clusters of up to 200 particles.
The simple and generic pair interaction model of binary mixtures combines the repulsion of monodisperse cores with short range attractions of three different magnitudes.
The interaction parameters were set to mimic a combination of gold, silver and polymeric nanoparticles in hexane, but the main results hold more generally for mixtures of particles with different van der Waals interactions, and even metal vapors. 

We analyzed several quantities that characterize the structure of the clusters and discussed the factors that determine the interplay between the packing of the spherical particles (entropy) and the optimization of the number of energetically favorable neighbors in the clusters (energy).
By global optimization we predict structures that could be observed in experiments, in particular, we predict which compositions lead to core-shell clusters, Janus clusters and clusters with specific valences. The latter are based on icosahedral symmetry where the minority species is located at the vertices or on the planes.
If functionalized suitably, these clusters could be promising building blocks for colloidal molecules and crystals.
All clusters can be visualized at http://softmattertheory.lu/clusters.html

\begin{acknowledgments}
  This project has been financially supported by the National Research Fund (FNR) within the INTER-DFG project Agglo. Computer simulations presented in this paper were carried out using the HPC facilities of the University of Luxembourg.
\end{acknowledgments}

\bibliography{main}

\end{document}